\documentstyle[12pt]{article}
\title{Change of the Chern number at band crossings}
\author{J. Bellissard\thanks
{e-mail: jeanbel@irsamc2.ups-tlse.fr}\\  Universit\'e Paul Sabatier,
Toulouse, France \thanks {URA 505, CNRS  and Laboratoire de
Physique Quantique,  118, Route de Narbonne, 31062-Toulouse Cedex,
France}} 
\date{ }

\newtheorem{theo}{Theorem}

\newcommand{\bed}{\begin{description}}
\newcommand{\ed}{\end{description}}

\newcommand{\Ch}{{\rm Ch}}
\newcommand{\TR}{{\rm Tr}}

\newcommand{\Cc}{{\cal C}}

\newcommand{\Kk}{{\cal K}}

\newcommand{\Oo}{{\cal O}}

\newcommand{\CC}{{\bf C}}

\newcommand{\RR}{{\bf R}}

\begin{document}

\maketitle

\begin{abstract}
Let $H(\epsilon, x)$ be a finite dimensional hermitian matrix
depending on the variable $x$ taking its values on a 2D manifold
and changing with $\epsilon$. If at $\epsilon=0$ two bands are
touching, we give a formula for the change of Chern number of these
bands as $\epsilon$ passes through zero. 
\end{abstract}

\vspace{1cm}

\section{Introduction}

In many problems of quantum physics, one has to compute the band
spectrum of some Hamiltonian. This is what happens in solid
state physics for the electronic band spectrum. This is also the
case in molecular or nuclear physics with rotation bands in the
high spin limit \cite{Zh,LBe}. In most of these cases, the
Hamiltonian describing the dynamics can be reduced to a
$N$-dimensional hermitian matrix-valued continuous function $x\in
M\mapsto H(x)$, where $M$ is some manifold of parameters. For the
rotation bands, $M$ is the 2D sphere, whereas for electrons in a
perfect crystal, $M$ is a torus. Diagonalizing $H(x)$ for all $x$'s
gives rise to $x$-dependent eigenvalues $E_1(x),\ldots ,E_N(x)$
called the band functions. Moreover, each of the corresponding
eigenvectors define a line in $\CC^N$ depending continuously on $x$,
namely a line bundle. One important question to answer is whether
this bundle is trivial or not, namely if one can choose the
eigenvectors for each $x$ in such a way as to define a univalued
function over $M$. If $H(x)$ is real valued this is always
possible. However if $H(x)$ cannot be represented in some basis by
a real valued matrix depending continuously on $x$, this is not
possible in general. In physics such a situation happens
whenever the dynamics is not time-reversal invariant, for
instance if some magnetic field is present or if there are spin
couplings. The obstruction is measured by the Chern classes of the
line bundle. Such Chern classes have been related to the Quantum
Hall effect \cite{TKN,ASS,BES} for Bloch electrons in a uniform
magnetic field. For rotation bands, the occurrence of Chern numbers
gives rise to the quantization of a mechanical response to a
torque in molecules or nuclei \cite{Be95}. In the present work, we
will restrict ourself to the case for which the manifold $M$ is
two-dimensional, in view of its application to physics. 

We assume now that the Hamiltonian depends upon an additional real
parameter $\epsilon$ varying in a neighbourhood of
zero, say $[-1,+1]$. Since a famous result of Wigner and von Neumann
\cite{WV}, it is known that, if $H$ depends smoothly on $(x,\epsilon)$, the set
of points $(\epsilon ,x)$ for which two bands of $H(\epsilon, x)$ are touching
has codimension one generically. 

From now on, {\em we will assume that $M$ is an oriented smooth surface, namely it has
dimension $2$, and that we are in this generic situation}. This means that $H$
depends smoothly on $(x,\epsilon)$, and that there is a pair $E_- (\epsilon
,x)\leq E_+(\epsilon ,x)$ of neighbouring bands and a finite set $x_1,\ldots,
x_L$ of points in $M$ such that $E_- (\epsilon ,x)=E_+(\epsilon ,x)$ if and only
if $\epsilon =0$ and $x=x_j$ for some $j=1,\ldots,L$. The $x_i$'s are called the
``touching points'' of the two bands. We want to compute the change of the Chern
class of each of these bands while $\epsilon$ varies from $-1$ to $+1$. Such a
change has been observed in numerical computations \cite{AHT,AFL,Fau} related
to physical models.

To give the result in a precise form, we need some notation. First
of all, we will replace the $N\times N$ Hamiltonian $H(\epsilon,x)$
by an effective $2\times 2$ one describing the two bands under
study in a neighbourhood $\Oo$ of $\epsilon=0,x=x_j$. This is always
possible. Using the Pauli matrices $\vec{\sigma} =(\sigma_1,
\sigma_2, \sigma_3)$, this effective Hamiltonian can be written as 

\begin{equation}
\label{effHamil}
H_{eff}(\epsilon,x)=
	\frac{1}{2}(E_+(\epsilon,x)+E_-(\epsilon,x))
  +\vec{\sigma}\vec{f}(\epsilon,x)
\mbox{ , }
\end{equation}

\noindent where $\vec{f}$ takes on values in $\RR ^3$ and
$|\vec{f}|=(E_+(\epsilon,x)-E_-(\epsilon,x))/2$. From our
assumptions, it follows that $\vec{f}$ vanishes only at $(0,x_j)$
in $\Oo$. We then set $\vec{\omega}=\vec{f}/|\vec{f}|$ in the
complement of the $(0,x_j)$'s in $\Oo$. Given any surface $\Sigma_j$ in
$\Oo$ homotopic to a sphere centered at $(0,x_j)$ and not surrounding the other
touching points, we set

\begin{equation}
\label{Berryindex}
n(x_j)=
	\frac{1}{4\pi}
		\int_{\Sigma_j} 
			<\vec{\omega}|d\vec{\omega}\wedge d\vec{\omega}>
\mbox{ . }
\end{equation}

\noindent This is an integer which does not depends on the
choice of $\Sigma_j$, called the degree of the map
$\vec{\omega}$. This index is actually related to Berry's
phase. In his seminal paper \cite{Ber}, M. Berry showed that if at
$\epsilon =0$ the parameters $x$ turns around $x_j$ on a closed path
$\gamma$ close enough to $x_j$, the eigenvector $\psi_+$
corresponding to the $+$-band experiences a phase change given by

\begin{equation}
\label{Berryphase}
\Delta \phi =
 \frac{1}{2}	
  \int_{\Sigma} 
   <\vec{\omega}|d\vec{\omega}\wedge d\vec{\omega}>
\mbox{ , }
\end{equation}

\noindent where $\Sigma$ is any oriented surface in $\Oo$ with
boundary given by $\gamma$. For this reason, we propose to call
$n(x_j)$ the Berry index at $x_j$. Our main result is the following:

\begin{theo}
\label{deltaChern}
The difference $\Delta \Ch$ between the Chern
numbers of the subband corresponding to $E_+$ while $\epsilon$
varies from $-1$ to $+1$ is given by the sum of the Berry indices
at all touching points.
\end{theo} 

\noindent In Section~\ref{chap-index} we will compute the Berry index
in some cases. The generic situation corresponds to $\vec{f}$ having a non
vanishing Jacobian at the touching point and gives the following result:

\begin{theo}
\label{generic}
If $\vec{f}$ has a non vanishing Jacobian $J_{\vec{f}}(x_j)$ at the touching
point $(0,x_j)$, the Berry index is given by $n(x_j) =
J_{\vec{f}}(x_j)/|J_{\vec{f}}(x_j)|$, namely it is the sign of the Jacobian at
the touching point.  
\end{theo} 

\section{Proof of the main result}
\label{chap-proof}

\begin{figure}
\vspace{8cm}
\includegraphics{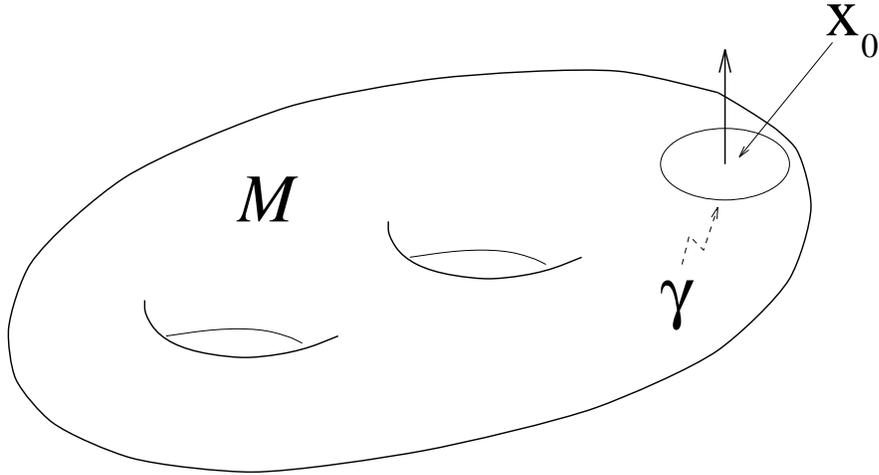}
\caption{\label{Msurf}{\sl The surface $M$}}
\end{figure}

\noindent In what follows, $M$ is an oriented smooth surface. Up to an
homeomorphism, it is given by the Poincar\'e construction from a polygon in the Poincar\'e disk (see
Fig.~\ref{Msurf}). Let us assume for simplicity that for $\epsilon \neq
0$, the spectrum of the Hamiltonian $H(\epsilon ,x)$ is simple. We
then denote by $P_l(\epsilon ,x)$ the eigenprojection associated
to the eigenvalue $E_l(\epsilon ,x)$. It is given either by 
$|\psi_l><\psi_l|$ where $\psi_l$ is a corresponding normalized
eigenvector, or by Cauchy's formula as

\begin{equation}
\label{projection}
P_l(\epsilon ,x)=	
	\frac{1}{2\imath \pi}
		\oint_{\Gamma}
			\frac{dz}{z-H(\epsilon ,x)}
\mbox{ , }
\end{equation}

\noindent where $\Gamma$ is a small circle centered at $E_l(\epsilon ,x)$ and
not surrounding the rest of the spectrum. Since the spectrum is simple for
$\epsilon \neq 0$, $P_l$ is smooth with respect to $(\epsilon ,x)$ in the
complement of the touching points in $[-1,+1]\times M$. For a given $\epsilon$
its second Chern class is given by the closed 2-form $\Omega_2$ on $M$ where

\begin{equation}
\label{omega2}
\Omega_2=
	\frac{1}{2\imath \pi}
		\TR (P_l dP_l \wedge dP_l)
\mbox{ . }
\end{equation}

\noindent The Chern number $\Ch(\epsilon)$ is obtained by integrating $\Omega_2$
over the surface $M$. That it is an integer is a standard result
\cite{LM,Hir,TKN}. Moreover, this number does not change with $\epsilon$ as long
as the $l$-th band remains isolated from the rest of the spectrum. Finally if
$P_l$ is two-dimensional, namely of the form $(1+\vec{\sigma}\vec{\omega})/2$
for some unit vector ${\omega}$ in $\RR ^3$, the two-form becomes $\Omega_2
=(\imath /2)<\vec{\omega}|d\vec{\omega}\wedge d\vec{\omega}>$. 

\begin{figure}
\vspace{9cm}
\includegraphics{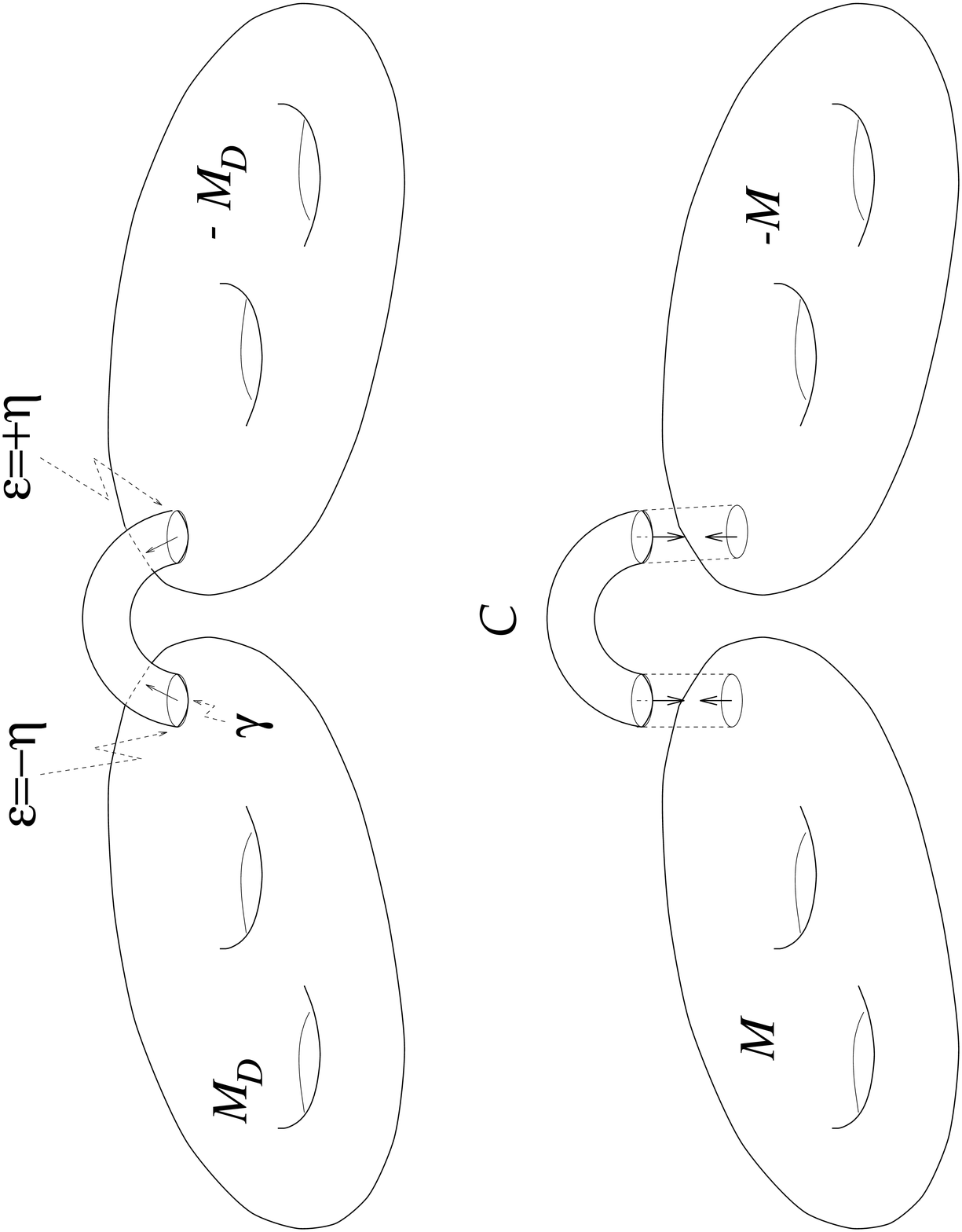}
\caption{\label{MDhat}{\sl The surface $\hat{M}_D(\eta)$}}
\end{figure}

Now, we assume that, at $\epsilon =0$, the band $E_+ =E_l$ touch
$E_- =E_{l-1}$ at one point $x_0\in M$ only. The general case can be
obtained in a similar way. Let $D$ be a small open neighbourhood of
$x_0$ in $M$ diffeomorphic to the unit disk in $\RR^2$. Let then
$\gamma$ be the oriented boundary of $D$ and $M_D$ be the complement
of $D$ in $M$ endowed with the same orientation as $M$. Therefore, as
an oriented path $\partial M_D =-\gamma$. For $0\leq \eta \leq 1$, we
construct a new surface $\hat{M}_{D}(\eta)$ by gluing together $\{
+\eta\} \times (-M_{D})$, $[-\eta,+\eta]\times \gamma$ and $\{ -\eta\}
\times (M_{D})$, where $-M'$ denotes the manifold $M'$ with opposite
orientation (see Fig.~\ref{MDhat}).

We give $\hat{M}_D(\eta)$ the natural orientation to
make it closed. As $\eta \rightarrow 0$ this surface is
homotopic to itself and becomes two copies of $M_D$ with
opposite orientations glued along $\gamma$. Moreover, for every
$\eta$, this surface contains no point on which the spectrum
of $H$ is degenerate. Therefore $P_l$ defines a line bundle on
it. Its Chern class is given by integrating $\Omega_2$ on
$\hat{M}_D(\eta)$. Owing to the homotopy invariance of this
Chern number, it is enough to compute it for $\eta=0$ and gives
the obvious result 

$$
\int_{\hat{M}_D(\eta)}\Omega_2 = 0
\mbox{ . }
$$

\noindent On the other hand, $\hat{M}_D(\eta)$ can be decomposed into three
closed surfaces (see Fig.~\ref{MDhat}), namely it is homologous to the sum of
the following three cycles $M_+=\{ +\eta\} \times(-M)$, $M_-=\{ -\eta \} \times
(M)$ and $\Cc$ which is obtained by gluing together $\{ +\eta\} \times(D)$, $\{
-\eta\} \times(-D)$ and $[-\eta,+\eta]\times \gamma$ with the corresponding
orientation. Integrating $\Omega_2$ over $M_+$ gives $-\Ch(+1)$, namely the
opposite of the Chern number of the $l$-th band for $\epsilon >0$. Integrating
it over $M_-$ gives $\Ch(-1)$, namely the  Chern number of the $l$-th band for
$\epsilon <0$. Finally the contribution of $\Cc$ is precisely the Berry index
defined by eq.~(\ref{Berryindex}) in the introduction, because $\Cc$ is
homotopic to a sphere surrounding the touching point once.
\hfill $\Box$

\section{Computation of the Berry index}
\label{chap-index}

\paragraph{Generic touchings -}

In the generic case, the function $\vec{f}$, in eq.~(\ref{effHamil}),
expressed in a suitable local chart, admits a Taylor expansion about
the touching point $x_0$ given by

$$\vec{f}(\epsilon , x_0 +\xi) =
			\epsilon \vec{f}_0 +\xi_1 \vec{f}_1 + \xi_2 \vec{f}_2
				+ \Oo (\epsilon^2+\xi_1^2+\xi_2^2)
\mbox{ , }
$$

\noindent where $\xi =(\xi_1,\xi_2)$ is a small vector in $\RR^2$ and
the three vectors $\vec{f}_j$ are linearly independent. We do not
change the homology class of the closed two-form $\Omega_2$ in
eq.~(\ref{omega2}) by modifying $\vec{f}$ homotopically. We can
therefore ignore the remainder and deform continuously the basis
$\vec{f}_j$ to make it coincide with the canonical one up to
orientation. The orientation is given by the sign $\varepsilon$ of
$<\vec{f}_0| \vec{f}_1\times \vec{f}_2>$, namely the sign of the
Jacobian of $\vec{f}$ at the touching point $\epsilon =0,x=x_0$.
Therefore, up to homotopy, we can assume $\vec{f}(\epsilon ,x_0+\xi) =
\varepsilon (\epsilon, \xi)$. Using polar coordinates $(\theta ,\phi)$
in $\RR^3$, this gives $\vec{\omega}=\vec{f}/|\vec{f}|=\varepsilon
(\cos(\theta),\sin(\theta)\cos(\phi), sin(\theta)\sin(\phi))$, and thus
$\Omega_2=\varepsilon \sin(\theta) d\theta \wedge d\phi$. Since $\Cc$
is homotopic to a sphere, it follows that the Berry index in this case
is 

$$
n(x_0) = \varepsilon= 
 \mbox{\rm sgn} \; \; \det \left (
  \frac{\partial \vec{f}}{\partial (\epsilon,x)}
   \right )_{\epsilon =0,x=x_0}
    = \frac{J_{\vec{f}}(x_0)}{|J_{\vec{f}}(x_0)|}
\mbox{ . }
$$

\hfill $\Box$

\paragraph{Parabolic touchings -}

Parabolic touching may occur non generically. Such examples have been
encountered in the literature \cite{AFL,Fau,Arm} in connection with
either quantum chaos or Bloch electrons in magnetic fields.
By redefining the parameter $\epsilon$ if necessary, we may assume that
$\vec{f}$ vanishes linearly in $\epsilon$ at $x=x_0$. 
We will say that the touching is ``parabolic non degenerate'' whenever 

$$\vec{f}(\epsilon, x_0+\xi)=
   \epsilon \vec{f}_0 +
				\xi_1^2\vec{f}_{11} +
					\xi_2^2\vec{f}_{22}+
						2\xi_1\xi_2 \vec{f}_{12}+
							\Oo (\epsilon ^2, |\xi|^3)
\mbox{ , }
$$

\noindent where the three vectors
$\{\vec{f}_{11},\vec{f}_{22},\vec{f}_{12} \}$ are linearly
independent. Let $\varepsilon =\pm 1$ be the orientation of this basis.
Again by homotopy, we can neglect the remainder and replace these three
vectors by the canonical basis of $\RR^3$, up to the sign
$\varepsilon$. Thus $\vec{f}=\epsilon \vec{f}_0+\varepsilon
\vec{g}(\xi)$ where

$$\vec{g}(\xi)= (\xi_1^2,\xi_2^2,2\xi_1 \xi_2)
\mbox{ . }
$$

\noindent Whenever $\xi$ varies in $\RR^2$, $\vec{g}(\xi)$ describes
the cone $\Kk = \{ (x,y,z)\in \RR^3; x\geq 0, y\geq 0,
(x+y)^2=(x-y)^2 + z^2\}$. This is a cone with vertex at the origin,
with axis parallel to $(1,1,0)$ and with basis given by the ellipse
perpendicular to the axis. If moreover, $\xi$ describes a small circle
$\gamma$ in $\RR^2$, $\vec{g}(\xi)$ describes an ellipse similar to
the basis of the cone and approaching the cone vertex as the
radius of $\gamma$ decreases to zero. In addition, this ellipse is
described twice as $\gamma$ is described once. Therefore we find two
situations compatible with our hypothesis. Either $\pm \vec{f}_0$ belongs
to the interior of this cone, and then the surface $\vec{f}(\Cc)$
surrounds the origin twice, with the orientation given by
$\varepsilon$, or it does not and $\vec{f}(\Cc)$ does not surround the
origin at all. In the former case the Berry index is
$n(x_0)=2\varepsilon$ in the latter $n(x_0)=0$. 

If the touching is ``parabolic degenerate'', namely if the three
vectors $\{\vec{f}_{11},\vec{f}_{22},\vec{f}_{12} \}$ are linearly
dependent, the same argument shows that the cone $\Kk$ is flat so that
the Berry index vanishes. This is the case if for instance
$\vec{f}_{12}=0$ \cite{Arm}.


\newpage

\begin{thebibliography}{9999}

\bibitem{Zh}V.D. Pavlov-Verevkin, D.A. Sadovskii, B.I. Zhilinsky, ``On the
Dynamical Meaning of the Diabolical Points'', {\sl Europhys. Lett.}, {\bf 6},
(1988), 573-578.

\bibitem{LBe}C. Lauzeral, J. Bellissard, {\sl in preparation}.

\bibitem{TKN} D. Thouless, M. Kohmoto, M. Nightingale et M. den Nijs,
``Quantized Hall conductance in two-dimensional periodic potential'', {\sl Phys.
Rev. Lett.}, {\bf 49}, (1982), 405-408.

\bibitem{ASS}B. Simon,``Holonomy, the quantum adiabatic theorem and Berry's
phase'', {\sl Phys. Rev. Lett}, {\bf 51}, (1983), 2167-2170.
51.

\bibitem{BES}J. Bellissard, A. van Elst, H. Schulz-Baldes,``The Non Commutative
Geometry of the Quantum Hall Effect'', {\sl J. Math. Phys.}, 35, (1994),
5373-5471, and references therein.

\bibitem{Be95}J. Bellissard, {\sl in preparation}.

\bibitem{WV}J. von Neumann, E.P. Wigner, {\sl Phys. Z.}, {\bf 30}, (1929),
467-470; E. Teller, {\sl J. Chem. Phys.}, {\bf 41}, (1937), 109-116.

\bibitem{AHT}Z. Tesanovic, F. Axel, B. Halperin, ``Hall crystal versus Wigner
crystal'', {\sl Phys. Rev.}, {\bf B39}, (1989), 8525-8551.

\bibitem{AFL}P. Leb\oe uf, J. Kurchan, M. Feingold, D.P. Arovas, ``Topological
aspects of quantum chaos'', {\sl Phys. Rev. Lett.}, {\bf 65}, (1990), 3076; {\sl
Chaos}, {\bf 2}, (1992), 125.

\bibitem{Fau}F. Faure, ``Approche g\'eom\'etrique de la limite semi-classique
par les \'etats coh\'erents et m\'ecanique quantique sur le tore'', Th\`ese,
Univ. J. Fourier, Grenoble, (1993).

\bibitem{Ber}M.V. Berry, ``Quantal phase factors accompanying adiabatic
changes'', {\sl Proc.R. Soc. Lond.}, {\bf A392}, (1984), 45-57.

\bibitem{LM}
H. Lawson, M. Michelson, 
{\sl Spin Geometry},
(Princeton University Press, Princeton, 1989).

\bibitem{Hir}
F. Hirzebruch, 
{\sl Topological Methods in algebraic geometry},
(Springer-Verlag, Berlin, 1966).

\bibitem{Arm}A. Barelli, R. Fleckinger, ``Semi-classical analysis of Harper-like
models'', {\sl Phys. Rev.}, {\bf B46}, (1992), 11559-11569.

\end{thebibliography}
\end{document}